\documentclass{jps-cp}
\usepackage{txfonts} 
\usepackage{color}
\usepackage{verbatim}

\newcommand{\bsigma}{\boldsymbol{\sigma}}

\newcommand{\eps}{\varepsilon}
\newcommand{\cepd}{Ce$_3$Bi$_4$Pd$_3$}

\def\ra{\rangle}
\def\la{\langle}
\def\up{\uparrow}
\def\dn{\downarrow}
\def\Hc{{\rm H.c.}}
\def\k{{\bf k}}

\title{Weyl-Kondo Semimetal: Towards Control of Weyl Nodes}

\author{Sarah E. \textsc{Grefe}$^{1}$, Hsin-Hua \textsc{Lai}$^1$, Silke \textsc{Paschen}$^2$, and Qimiao \textsc{Si}$^{1}$}
\inst{$^{1}$Department of Physics and Astronomy \& Rice Center for Quantum Materials, Rice University, Houston, Texas 77005, USA\\
$^{2}$Institute of Solid State Physics, Vienna University of Technology, Wiedner Hauptstra{\ss}e 8-10, 1040 Vienna, Austria}
\email{seg5@rice.edu}
\recdate{September 17, 2019}

\abst{Heavy fermion semimetals represent a promising setting to explore topological metals driven by strong correlations. In this paper, we i) summarize the theoretical results in
 a Weyl-Kondo semimetal phase for a strongly correlated model with inversion-symmetry-breaking and time-reversal invariance, and the concurrent work that has experimentally discovered this phase in the non-magnetic non-centrosymmetric heavy fermion system Ce$_3$Bi$_4$Pd$_3$; and ii) describe what is expected theoretically when the time-reversal symmetry is also broken.}

\kword{Topological phases, topological semimetals, symmetry protected topological phases, Weyl semimetals, Kondo lattice systems, Anderson lattice model, diamond lattice}

\begin{document}
\maketitle

\section{Introduction}\label{sec:intro}

The usual focus in the study of topological states of electronic matter has been on insulating states, a tradition that dates back to the heyday of quantum Hall systems and continued into the era of topological insulators. 
Topological metals have only been studied recently~\cite{RevModPhys.90.015001}. 
An outstanding question is how topological metals can arise as a result of strong correlations. 
Of particular interest in this context are heavy-fermion semimetals~\cite{SiPaschen2013}, in which the electron correlations are strong and spin-orbit coupling may be large.
Recently, concurrent theoretical and experimental studies have advanced the notion of a Weyl-Kondo semimetal. 
In Ref.~\cite{Grefe_PNAS_2017,WKSM_extended}, a theoretical analysis was carried out for a strongly correlated model in which the inversion symmetry (IS) is broken while the time-reversal symmetry (TRS) is preserved. 
Meanwhile, Refs.~\cite{Dzsaber2016,2018arXiv181102819D,dzsaber2019quenching} have experimentally discovered this phase in the non-magnetic non-centrosymmetric heavy fermion system \cepd. 

In this paper, we briefly summarize these developments in Sec.~\ref{sec:wksm+trs}, and describe what is expected theoretically when the TRS is also broken in Sec.~\ref{sec:trs+is-breaking}. 
Our results suggest that, upon the tuning of parameters that break both the TRS and IS, Weyl semimetal (WSM) nodes are moved about and annihilated through critical phase boundaries with distinct quadratic-band touchings. 
In Sec.~\ref{sec:discussion}, we discuss these results and point to further studies for the future.

\section{Weyl-Kondo semimetal with time-reversal invariance}\label{sec:wksm+trs}

\begin{figure}[tbh]
	\includegraphics[width=2in]{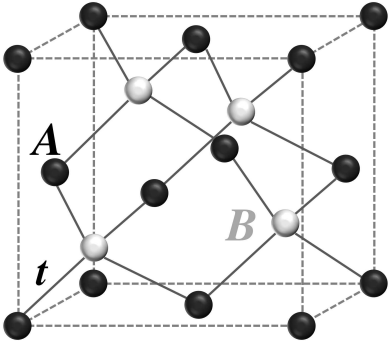}
\caption{The unit cell of the diamond lattice showing the two fcc sublattices $A,~B$ and the fermion hopping amplitude $t$ along the nearest-neighbor bonds.
			}
\label{fig:lattice}
\end{figure}

In Ref~\cite{Grefe_PNAS_2017,WKSM_extended}, we proposed a periodic Anderson model on the diamond lattice (Fig.~\ref{fig:lattice}) of the form
\begin{align}
\mathcal{H}
_{\mathrm{PAM}}
 &= \mathcal{H}_c + \mathcal{H}_d + \mathcal{H}_{cd},\label{eq:hamiltonian}
\\
\mathcal{H}_c &= t\sum_{\la i j \ra, \sigma } \left( c_{i \sigma}^\dagger c_{j \sigma}+ \Hc \right)-\mu \sum_{i, \sigma} n_{i \sigma}^c
	+i\lambda \sum_{\la\la i j \ra\ra} \left[ c^\dagger_{i \sigma} \left( {\bsigma}\cdot {\bf e}_{ij} \right) c_{j\sigma} - \Hc\right]
	+ m \sum_{j,\sigma} (-1)^j c^\dagger_{j\sigma} c_{j\sigma},\label{eq:Hc}\\
\mathcal{H}_d &= E_d\sum_{i, \sigma} n_{i \sigma}^d
	+ \mathcal{U} \sum_i n^d_{i \up} n^d_{i \dn}\label{eq:Hd},
\\
\mathcal{H}_{cd} &= V\sum_{i,\sigma} (c^\dagger_{i\sigma} d_{i\sigma} + \Hc ).\label{eq:Hcd}
\end{align}
It has the usual components of an Anderson lattice model.
The $d$-operators describe the physical $4f$-electrons; their highly localized wavefunctions can be well treated in an atomic limit, with an energy level $E_d$ and a mutual Coulomb interaction that assigns an energy penalty $\mathcal{U}$ to
double occupation in Eq.~\ref{eq:Hd}.
The $c$-operators represent the $spd$-conduction electrons in the system, albeit with spin-orbit coupling ($\lambda$) and a sublattice-dependent term $m$ which breaks IS.
Finally there is uniform on-site hybridization between the localized and free conduction electrons of Eq.~\ref{eq:Hcd}, which is in the Kondo limit for large $\mathcal{U}$,
 with a Kondo coupling $J_K \sim V^2/|E_d|$.

We studied the model with $n_d+n_c=1$ per site, corresponding to a quarter filling.
To analyze the model in the strong-coupling limit ($\mathcal{U}/t\rightarrow\infty$), we implemented an auxiliary boson 
representation to effectively treat the Coulomb repulsion as fixing the localized electrons' density.
When the Kondo coupling vanishes
(via the hybridization $V\rightarrow0$), an electron occupies the $d$-level,
forming a Mott insulator,
and the conduction electron bands are empty, leading to a topologically trivial state. 
In the Kondo-screened state 
(for $V$ above a threshold value $V_c$),
however, strongly renormalized Weyl nodes emerge and are pinned at the Fermi energy.
This shows that the localized electron species play a central role in the formation of the Weyl nodes, which have a narrow bandwidth on the order of the Kondo temperature ($T_K$), a nodal velocity ($v^*$) that is reduced from a normal metal value ($v$) by several orders of magnitude, and the strong coupling from the localized electrons renormalizes the bands to fulfill the density constraint.

These characteristics are manifested in heavy fermion semimetals through several thermodynamic and transport signatures.
The specific heat behaves as 
$c_v=\Gamma
\, T^3$
with $\Gamma\propto(v/v^*)^3$ at an onset temperature 
of $T_K$ (well-below the Debye temperature for phonons).
This result is robust to residual interactions
(Ref.~\cite{Grefe_PNAS_2017}, supporting information).
This behavior was observed in specific heat measurements on \cepd~\cite{Dzsaber2016}.
In Ref.~\cite{2018arXiv181102819D}, \cepd~ was found to have a giant spontaneous Hall effect in the nonmagnetic phase without application of a magnetic field (TRS is preserved in the absence of a driving electric field), and a Hall resistivity that is even with respect to the magnetic field.
Thus there has been substantial progess in studying heavy fermion semimetals which preserve TRS in non-centrosymmetric systems.
Related observations have been reported in YbPtBi~\cite{Guo:2018aa}.

The IS-broken Weyl semimetal phases are generally well studied and understood as arising from a combination of crystalline and non-spatial symmetries, and has thus far been a major focus on studies of topological materials.
We now shift our attention to topological semimetal phases that are not protected by TRS.

\section{The effect of breaking both time-reversal and inversion symmetries}\label{sec:trs+is-breaking}

Recent experimental studies~\cite{dzsaber2019quenching}
have used a magnetic field to tune and eventually annihilate the Weyl nodes of Ce$_3$Bi$_4$Pd$_3$.
We are interested in understanding the behavior of 
this material, especially how
the TRS breaking term controls its Weyl nodes.
Since this issue is completely open, we will start by focusing on the symmetry aspects of the problem, namely how the Weyl nodes
evolve from the Brillouin zone (BZ) boundary
to its interior and their eventual annihilation, as the degrees of IS and TRS breaking are tuned.
As a first step in the understanding, we will focus on the single fermion flavor model.
Because the space-group symmetry is expected to play an important role in this evolution,
we expect that the qualitative aspect of our results on the control of the Weyl nodes by the IS and TRS breaking potentials 
will be relevant to the full Kondo lattice model.
%

Thus,
we consider the model $\mathcal{H}_c$ in the presence of a Zeeman coupling. 
The effect of breaking TRS but preserving the IS was reported in Ref.~\cite{Grefe_PNAS_2017}
(supporting information).
There it was found that a Weyl semimetal phase evolves in two stages, with nodes occurring 
in the BZ interior instead of the BZ boundary.

Here, we study a variation on the conduction electron-only model by including terms that break both IS and TRS.  
We tune the symmetry-breaking terms to map out a phase diagram and find three distinct WSM phases, separated by critical phase boundary lines of distinct quadratic-band touching phases.

We consider $\mathcal{H}_c$ defined earlier on the diamond lattice in Fig.~\ref{fig:lattice}, with the addition of a Zeeman coupling:
\begin{align}
\mathcal{H} &= \mathcal{H}_c + M_z \sum_i \hat{z} \cdot \left( c^\dagger_{i\sigma}  \bsigma c_{i\sigma} \right).\label{eq:ham}
\end{align}
As already specified in Eq.~\ref{eq:Hc}, the onsite potential $+m(-m)$ on the $A(B)$ sublattice
breaks IS on the diamond lattice
and reduces
the space group from \#227 ($Fd\bar{3}m$) to \#216 ($F\bar{4}3m$, zincblende).
The Zeeman term tuned by $M_z$
 breaks TRS.
In this work, we fix $\lambda=1/2$ to facilitate band inversion as a necessary (but insufficient) condition towards a topological semimetal state.

The eigenenergies are obtained by diagonalizing the Bloch Hamiltonian.
Although the full dispersion is analytically tractable, it is simpler to look at the eigenenergies when $M_z=0,~m\neq0$ and vice versa.
These are:
\begin{align}
	\eps^\alpha_\beta(M_z=0)&= \alpha\sqrt{u_1^2+u_2^2+ (m+\beta \lambda D_\k)^2 },\\
	\eps^\alpha_\beta(m=0)&= \alpha\sqrt{u_1^2+u_2^2+\lambda^2 D^2_\k +M_z^2 +2\beta M_z\sqrt{ u_1^2+u_2^2+\lambda^2 D_z^2 }},
\end{align}
where $\alpha,\beta=\pm1$ index the four resulting bands, and $D_\k=\sqrt{{\bf D}(\k) \cdot{\bf D}(\k)}=\sqrt{D_x^2+D_y^2+D_z^2}$. 
The notations are described in detail in Ref~\cite{Grefe_PNAS_2017}.
We immediately see that $m$ competes with the spin orbit coupling term, whereas $M_z$ competes with the $z-$component of the spin orbit interaction and the $u_1,~u_2$ terms 
that
come from the hopping terms proportional to $t$.

\subsection{Phase diagram: TRS- and IS-breaking coexistence and competition}\label{sec:phases}

\begin{figure}[tbh]
	\includegraphics[width=3in]{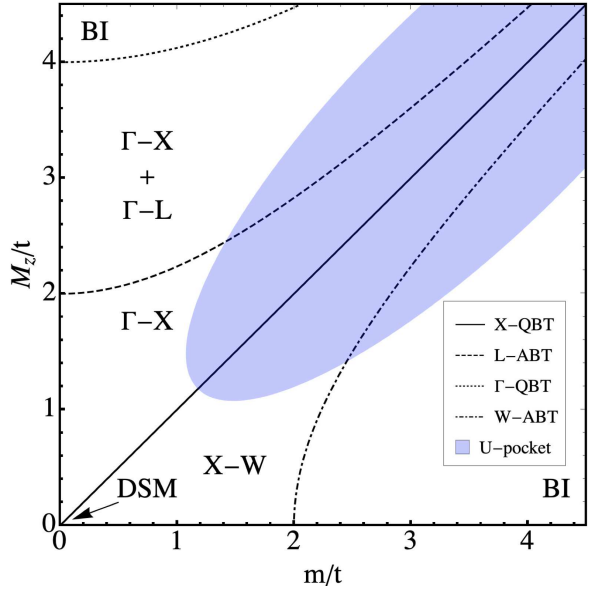}
		\caption{Phase diagram for the WSM phases that our model captures with breaking both time-reversal and inversion symmetries.
		DSM=Dirac semimetal, BI=Band insulator, $X/\Gamma$-QBT = quadratic band touching at $X$ or $\Gamma$, $L/W$-ABT=anisotropic band touching at $L/W$, and $U$-pocket denotes an approximate region where a Fermi pocket at $U$ emerges and coexists with the labeled phases.
		The regions with labels of high symmetry lines are Weyl semimetals with the nodes found along those lines.
		}
\label{fig:phases}
\end{figure}

\begin{figure}[h]
	\includegraphics[height=2.2in]{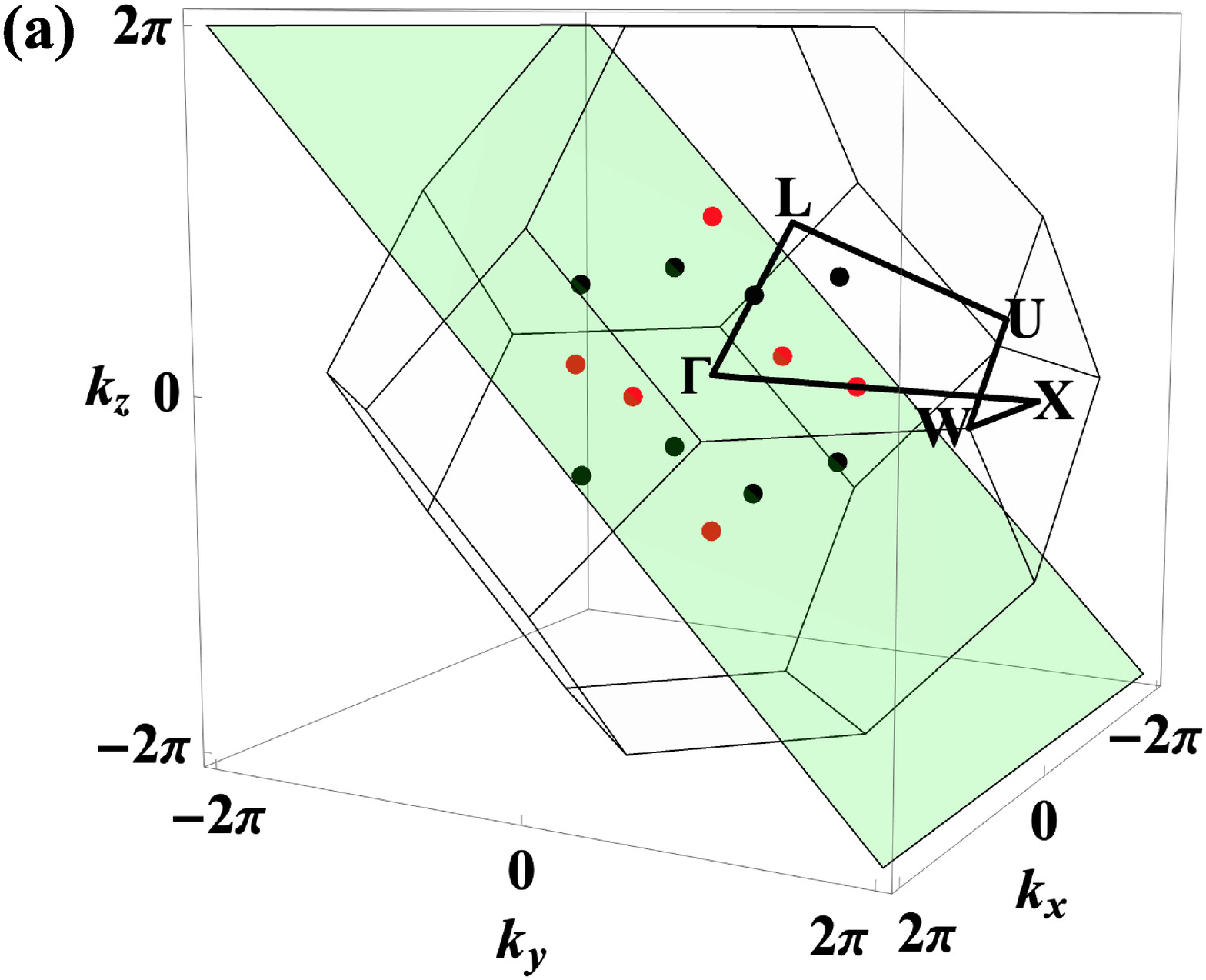}
	\includegraphics[height=2.15in]{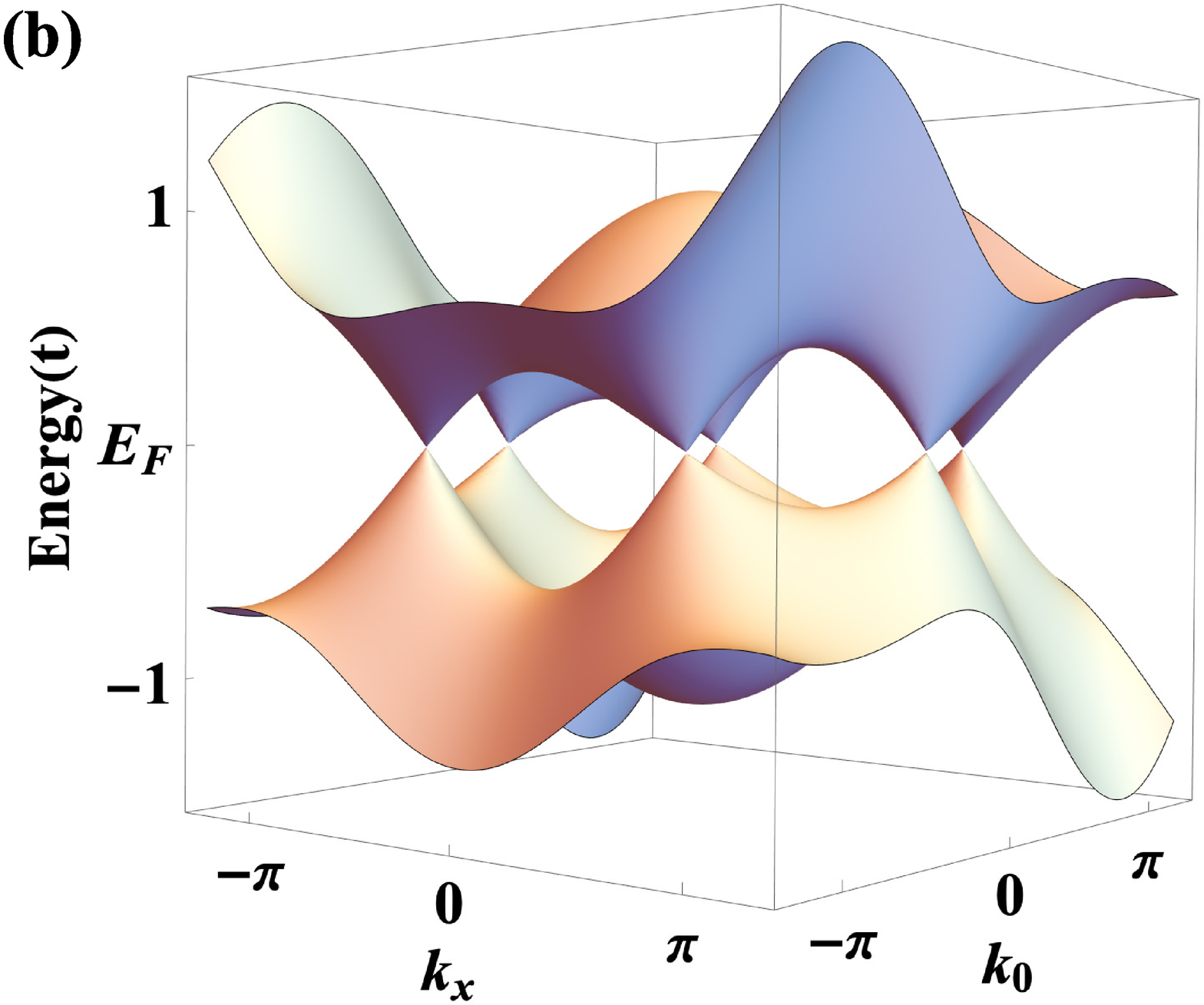}
	\caption{\label{fig:gamXL-bz-nodes}
	(a) Locations in the Brillouin zone where Weyl nodes appear for $M_z=3,~m=1$. 
High symmetry points are labeled, and Weyl node colors are red for $\Gamma-X$, black for $\Gamma-L$.
The $[011]$ plane that intersects $\Gamma-X$ and $\Gamma-L$ nodes is shown in green.
(b) Eigenenergy dispersion projected onto the $[011]$ plane shown in green in (a) that intersects with the $\Gamma-X$ and $\Gamma-L$ nodes.
}

\end{figure}

We now describe the phases we encounter as we tune $M_z,m\geq0$, as shown in Fig.~\ref{fig:phases}.
Note that these phases are symmetric with respect to the signs of $M_z$ and $m$.

When TRS and IS are preserved ($M_z=m=0$), this model realizes a Dirac semimetal, labeled at the origin of the phase diagram.
In this phase, fourfold degenerate Dirac nodes are protected by non-symmorphic symmetries at the $X$-points $\k_X=2\pi\hat{r}$, 
which are time-reversal invariant momenta (TRIM).
The Dirac point represents a critical point of the Fu-Kane-Mele model's phase diagram of topological insulators~\cite{FKMmodel07}, which has been extended to WSM phases through IS breaking terms~\cite{PhysRevB.78.165313,Ojanen13}.

Generically, the addition of TRS or IS breaking terms will split Dirac nodal points into one or more pairs of Weyl nodes, which is precisely what happens in our model.
The next subsections explain the WSM phases and the boundaries between them in terms of which high symmetry lines contain Weyl nodes or quadratic band touching points.
In Fig.~\ref{fig:gamXL-bz-nodes}(a) we show the fcc BZ with its high symmetry points and node locations labeled for an example phase, as well as the $[011]$ plane along which we plot the dispersion of the $\Gamma-X+\Gamma-L$ Weyl semimetal (Sec.~\ref{sec:gxglWSM}) in Fig.~\ref{fig:gamXL-bz-nodes}(b).

\subsubsection{$X-W$ Weyl semimetal}

The $M_z=m>0$ phase boundary (Fig.~\ref{fig:phases} solid line) marks a massful semimetal with a fourfold-degenerate quadratic band touching point at $X$, labeled $X-QBT$.
When $M_z<m$ and $m$ is small, the quadratic bands split into two pairs of Weyl nodes, and move along the $X-W$ lines, coexisting with the Zeeman field as long as $\sqrt{m^2-4}<M_z<m$.
As $m$ increases, the Weyl nodes move towards the $W$ points, until the $W-ABT$ boundary (dashed-dotted line of bottom bound $M_z=\sqrt{m^2-4}$). 
Along this boundary, an anisotropic band touching occurs at $W$,
 with linear dispersion in the $W-U$ direction and quadratic along $X-W$, marking a critical line between semimetal and trivial~\cite{Ojanen13} band insulator ($BI$).

\subsubsection{$\Gamma-X$ Weyl semimetal}

When $M_z>m\neq0$, the $X-QBT$ bands split into Weyl nodes along $\Gamma-X$.
Here a WSM phase distinct from the $X-W$ WSM phase emerges, with the restriction that $m<M_z<\sqrt{m^2+4}$.
Because there are six $\Gamma-X$ lines, this produces three pairs of nodal points in the BZ.

The upper bound $M_z=\sqrt{m^2+4}$ marks the $L-ABT$ phase boundary marking an anisotropic band touching at the $L$ point, dotted line in Fig.~\ref{fig:phases}).
Here, a quadratic touching along the $\Gamma-L$ line meets a linear dispersion along $L-U$.

\subsubsection{\label{sec:gxglWSM} $\Gamma-X+\Gamma-L$ Weyl semimetal}

As one moves in parameter space above the $L-ABT$ line, new Weyl points evolve from the critical touching at $L$, where linearly dispersing nodes travel along the $\Gamma-L$ line.  
These new nodes coexist with the $\Gamma-X$ nodes, and their semimetal phase persists as long as $M_z$ is between $\sqrt{m^2+4}<M_z<\sqrt{m^2+16}$.
The eigenenergies of this phase are plotted in Fig.~\ref{fig:gamXL-bz-nodes}(b), in the $[011]$ plane, where one can see both sets of nodes along $\Gamma-X$ and $\Gamma-L$.

The $\Gamma-QBT$ phase boundary ($M_z=\sqrt{m^2+16}$) is marked by a dotted line in Fig.~\ref{fig:phases}.
Along this boundary, the nodes from $\Gamma-X$ lines and $\Gamma-L$ lines all meet at $\Gamma$ and become a single, twofold degenerate quadratic band touching, labeled $\Gamma-QBT$.
As the Zeeman field is increased, a band gap opens, and the system becomes an insulator.

\subsubsection{$U-$Fermi surface coexistence metallic phase}

Finally, a last phase diagram feature warrants analysis, namely the ``$U$-pocket'' region approximately represented in blue.
When both the Zeeman splitting and the sublattice symmetry breaking are larger and close enough to each other, a Fermi pocket gradually emerges as a Fermi surface at $U$ as $M_z\sim m$ increases.
This metallic pocket does not eliminate most nodal point features or phase boundaries, except for those close to the $L-ABT$ boundary for large values of $M_z\sim4$.

\section{Discussion and Summary}\label{sec:discussion}

Our main result is the phase diagram shown in Fig.~\ref{fig:phases} as a function of the Zeeman coupling $M_z$ and the IS-breaking potential $m$.
Within our model, the Weyl semimetal phases survive the combined breaking of TRS and IS, and when the Zeeman term and the sublattice potential are larger and similar in size, a Lifshitz transition occurs where a hole pocket emerges at the $U$ point of the Brillouin zone.

The most important phase boundary is the $X-QBT$ line for balanced $M_z=m$, because it evolves from the critical point of the Fu-Kane-Mele model which serves as the multi-dimensional intersection of several nonlocal and space group symmetries.
For this reason, it represents a nexus of many phases.
In our work's two-dimensional slice of the phase diagram, the $X-QBT$ boundary marks a 
change from non-symmorphic symmetry-driven nodal degeneracy, enforced to exist on the BZ \emph{boundary} (e.g. the $X-W$ WSM), to where the TRS-breaking terms dominate and nodal phases evolve along high symmetry lines within the \emph{interior} of the BZ.

The phase boundaries are critical lines where the Weyl nodes become quadratic band touching
at a high-symmetry point, just before the nodes annihilate or follow a different high symmetry line, depending on the neighboring phase.

In a full treatment of the Anderson lattice model, the $f$-electrons should be more responsive to the Zeeman field than the conduction electrons.
In turn, this is expected to directly connect with the 
recent experiments~\cite{dzsaber2019quenching} on the tuning of the Kondo-driven Weyl nodes by a magnetic field.

\section{Acknowledgements}
We thank J. Cano
and S. Dzsaber
 for illuminating discussions. This work has been supported by the NSF (DMR-1920740), 
 the Robert A. Welch Foundation (C-1411) and the ARO (W911NF-14-1-0525). 
Work in Vienna was supported by the Austrian Science Fund (DK W1243, I2535, I4047, and P29296).

\end{document}